\documentclass[preprint,12pt]{elsarticle}

\usepackage{bm}% bold math
\PassOptionsToPackage{hyphens}{url}
\usepackage{hyperref}
\usepackage{breakurl}
\usepackage{amsmath}
\usepackage{amssymb}

\begin{document}
\sloppy

% \preprint{APS/123-QED}
\begin{frontmatter}

\title{A fast multigrid-based electromagnetic eigensolver for
curved metal boundaries on the Yee mesh\tnoteref{funding}}

\tnotetext[funding]{This work was supported by the U.~S.~Department of
Energy grant DE-FG02-04ER41317.}

\author[colorado]{Carl A.~Bauer\corref{me}}
\cortext[me]{Corresponding author: carl.bauer@colorado.edu}
\ead{carl.bauer@colorado.edu}

\author[colorado]{Gregory R.~Werner}

\author[colorado,techx]{John R.~Cary}

\address[colorado]{Department of Physics and the Center for Integrated
Plasma Studies, University of Colorado, Boulder, Colorado 80309}
\address[techx]{Tech-X Corporation, Boulder, Colorado 80303}

%\author{Carl A.~Bauer$^{1,*}$}
%
%\author{Gregory R.~Werner$^1$}
%\author{John R.~Cary$^{1,2}$}
%
%\affiliation{$^1$Department of Physics and the Center for Integrated
%Plasma Studies, University of Colorado, Boulder, Colorado 80309}
%
%\affiliation{$^2$Tech-X Corporation, Boulder, Colorado 80303}
%\affiliation{$^*$Corresponding author: bauerca@colorado.edu}

%%%%%%%%%%%%%%%%%%%%%%%%%%%%%%%%%%%%%%%%%%%%%%%%%%%%%%%%%%%%%%%%%%%
\begin{abstract}

For embedded boundary electromagnetics using the
Dey--Mittra~\cite{dey1997lcf} algorithm, a special grad-div
matrix constructed in this work allows use of multigrid methods
for efficient inversion of Maxwell's curl-curl matrix.
Efficient curl-curl inversions are demonstrated within a
shift-and-invert Krylov-subspace eigensolver (open-sourced
at \url{https://github.com/bauerca/maxwell}) on the
spherical cavity and the 9-cell TESLA superconducting
accelerator cavity.  The accuracy of the Dey--Mittra
algorithm is also examined: frequencies converge with
second-order error, and surface fields are found to converge
with nearly second-order error.  In agreement with previous
work \cite{nieter2009application}, neglecting some
boundary-cut cell faces (as is required in the time domain
for numerical stability) reduces frequency convergence to
first-order and surface-field convergence to zeroth-order
(i.e.~surface fields do not converge).  Additionally and
importantly, neglecting faces can reduce accuracy by an
order of magnitude at low resolutions.

\end{abstract}


\end{frontmatter}

%\pacs{42.60.Da, 42.70.Qs, 41.75.Lx }

%\maketitle

%%%%%%%%%%%%%%%%%%%%%%%%%%%%%%%%%%%%%%%%%%%%%%%%%%%%%%%%%%%%%%%%%%%%

\section{Introduction}

The Dey-Mittra electromagnetics algorithm simulates smooth
curved per\-fectly-conducting boundaries using the Yee
finite-difference technique \cite{yee1966nsi,dey1997lcf}.
The algorithm is often called a cut-cell or
embedded-boundary technique since the mesh does not conform
to the geometry of the conducting boundary (grid cells,
faces, and edges are ``cut'' by boundaries). In the
time-domain, the Courant-Friedrichs-Lewy (CFL) condition
reduces the accuracy of the Dey--Mittra algorithm by
requiring the neglecting of some cut faces. More precisely,
the CFL condition states that the maximum stable timestep is
limited by the maximum eigenvalue (of the discretized
curl-curl matrix) and, in the Dey--Mittra algorithm, the
maximum eigenvalue can be inflated greatly by faces barely
cut by a boundary. A trade-off between accuracy and
wall-clock simulation time ensues; if fewer neglected faces
are desired (greater accuracy), the time-step must be
reduced \cite{dey1997lcf,nieter2009application}. In this
paper, we consider the Dey--Mittra algorithm in the
frequency-domain, where the CFL condition does not apply and
the full accuracy of the method can be used.

We begin by reviewing the two important aspects of the
problem: (1) the Dey--Mittra algorithm and (2) eigensolving
Maxwell's equations as discretized on the Yee mesh
(ultimately, this leads to the question: how does one invert
the curl-curl operator? Fortunately, this is well-studied
\cite{hiptmair1998multigrid,bochev2004improved,
clemens2005construction,arbenz2005multilevel,hu2006toward,
kolev2009parallel}). The advance of this paper is described
in Section \ref{sec:eigsolveDM}, and amounts to a
transformation of the discretized Dey--Mittra curl-curl
operator that allows efficient inversion by multigrid
techniques \cite{briggs2000multigrid}. Proof of performance
is given in the numerical results, where our eigensolver
attacks the spherical resonant cavity and the 9-cell TESLA
superconducting accelerator cavity. The code used throughout
this paper is open-sourced, and can be found at
\url{https://github.com/bauerca/maxwell}.

\section{The Dey--Mittra algorithm}

\begin{figure}[htbp]
\centering
\includegraphics[scale=0.4]{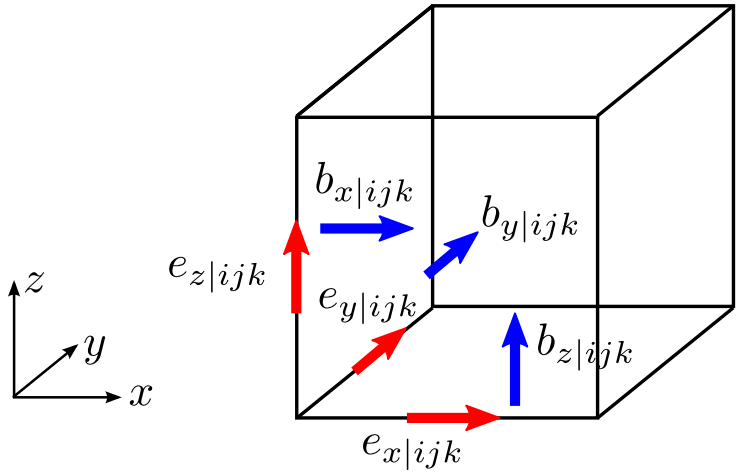}
\caption{Color online. Yee grid cell $ijk$. Electric (Magnetic) field
components are centered on edges (faces).}
\label{fig:yee}
\end{figure}

Electromagnetic cavity eigenmodes are solutions to Maxwell's wave
equation subject to perfectly conducting boundary conditions; a
magnetic eigenmode satisfies
\begin{align}
\nabla \times \nabla \times {\bf B} &= k^2 {\bf B} \quad \text{in
$\Omega$} \label{eq:waveb} \\
\mathbf{n} \cdot \mathbf{B} &= 0 \quad \text{on $\partial \Omega$}
\label{eq:normB}
\end{align}
where $\Omega$ is the cavity interior, $\partial \Omega$ is the
perfectly conducting boundary, $\mathbf{n}$ is the normal to the
boundary, and $k = \omega/c$, where $\omega$ is the resonant angular
frequency and $c$ is the speed of light. We discretize Maxwell's equations
with the finite-difference Yee algorithm \cite{yee1966nsi},
labeling
the grid electric and magnetic field components as $e_{\alpha|ijk}$
and $b_{\alpha|ijk}$, respectively, where $\alpha$ is one of $x$, $y$,
or $z$ and $i$, $j$, and $k$ are integer grid cell indices. Figure
\ref{fig:yee} shows the spatially staggered component layout of the Yee
scheme which ensures the first-order accuracy (second-order error) of the
discretized curl operators.
In matrix-vector form, where
$\mathbf{b}$ ($\mathbf{e}$) is the vector of all $b_{\alpha|ijk}$
($e_{\alpha|ijk}$) components, the discretized version of
Eq.~(\ref{eq:waveb}) in vacuum is written
\cite{taflove1995computational,werner2007sfa}
\begin{equation}
{\rm C C^T } {\bf b} = k^2 {\bf b}.
\label{eq:discWaveB}
\end{equation}
The Yee layout guarantees that the curl of the electric field is the
transpose of the curl of the magnetic field, resulting in the
symmetric matrix of Eq.~\ref{eq:discWaveB} (the curl-curl matrix is
also positive semi-definite, i.e. $k^2 \geq 0$).

The Dey-Mittra algorithm is a modification of the Yee algorithm which
simulates curved perfectly conducting boundaries in 3D with
second-order error \cite{dey1997lcf,nieter2009application}.
The algorithm is based on the finite
integral interpretation of the Yee algorithm
\cite{weiland1977discretization,clemens2001discrete} where, for
example, the Yee Faraday update for $b_{x|ijk}$ (in the frequency
domain) is written as
\begin{equation}
-i \omega b_{x|ijk} = \frac{1}{a_{x|ijk}} \left (
  l_{y|ijk} e_{y|ijk} - l_{y|ijk+1} e_{y|ijk+1} +
  l_{z|ij+1k} e_{z|ij+1k} - l_{z|ijk} e_{z|ijk}
  \right ),
\label{eq:farInt}
\end{equation}
which is a representation of Faraday's Law in integral form: $-
i \omega \int \mathbf{B} \cdot d\mathbf{a} = \oint \mathbf{E} \cdot
d\mathbf{l}$.  In the above, $l_{\alpha|ijk}$ is the length of the
edge of the Yee grid cell on which the component, $e_{\alpha|ijk}$, is
centered (see Fig.~\ref{fig:yee}). Similarly, $a_{\alpha|ijk}$ is the
area of the cell face on which the component, $b_{\alpha|ijk}$, is
centered. In vacuum, $l_{x|ijk} = \Delta x$, $l_{y|ijk} = \Delta y$,
and $a_{x|ijk} = \Delta y \Delta z$ such that Eq.~(\ref{eq:farInt})
reduces to the usual Yee finite difference expression.

\begin{figure}[htbp]
\centering
\includegraphics[scale=0.4]{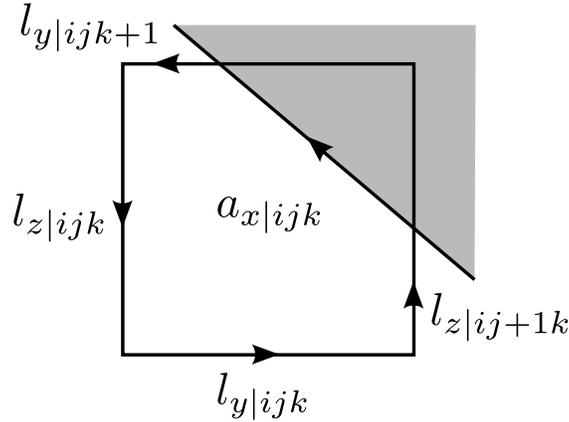}
\caption{An example of a cut face. The shaded region is conductor.
In the Dey--Mittra method, the loop integral for Faraday's update
is performed along the arrow-indicated contour. Because tangential
electric fields are zero at perfectly-conducting boundaries, the
line integral along the surface is skipped.}
\label{fig:dmFace}
\end{figure}

When a face, $a_{\alpha|ijk}$ is intersected by
a perfectly conducting boundary, the Dey--Mittra algorithm takes
$l_{\alpha|ijk}$ and $a_{\alpha|ijk}$ to be the portion of the
length and area, respectively, {\it outside} the conductor (see
Fig.~\ref{fig:dmFace}). This is a
physically meaningful
representation of Faraday's Law in integral form since the electric
field tangent to conducting boundaries vanishes. In matrix-vector
form, the Dey--Mittra algorithm changes Eq.~(\ref{eq:discWaveB}) to
\begin{equation}
{\rm A^{-1} C L C^T } {\bf b} = k^2 {\bf b}
\label{eq:discWaveBDM}
\end{equation}
where $\rm A$ is a diagonal matrix of cell face fractions (e.g.,
${\rm A}_{(x|ijk)(x|ijk)} = a_{x|ijk} / \Delta y \Delta z$)
and $\rm L$ is a diagonal matrix of cell edge fractions (e.g., ${\rm
L}_{(x|ijk)(x|ijk)} = l_{x|ijk} / \Delta x$).

Vanishing area fractions lead to very large elements in the inverse
area fraction matrix ${\rm A}^{-1}$ which can then
inflate the maximum eigenvalue of the Dey--Mittra curl-curl matrix
(relative to the vacuum Yee algorithm). For this reason, as
summarized in the next section, the Dey--Mittra algorithm
cannot be used to its full potential in the time-domain.
Following the next section, the rest of this paper considers
the frequency-domain, where this restriction is
no longer an issue.

\section{Dey--Mittra: weakened for the time-domain}

To use Dey--Mittra in the time-domain, one must neglect certain cut
faces with small area fractions, thus reducing the accuracy of the
algorithm.  In the standard Yee leap-frog time-stepping scheme, the
Courant-Friedrichs-Lewy (CFL) condition states that the maximum
eigenvalue of Eq.~\ref{eq:discWaveBDM} limits the maximum stable
timestep.
In 3D vacuum, the CFL condition on the timestep $\Delta t$ is
\begin{equation}
\Delta t < \Delta t_{\rm CFL} \equiv \frac{1}{c\sqrt{\frac{1}{\Delta
x^2} + \frac{1}{\Delta y^2} + \frac{1}{\Delta z^2}}}.
\end{equation}
In the Dey--Mittra algorithm, small area fractions amplify
elements of the matrix ${\rm A}^{-1}$. For some cuts, the edge lengths
in the numerator of Eq.~\ref{eq:farInt} do not compensate,
and the largest eigenvalue of
the Dey--Mittra wave equation (Eq.~\ref{eq:discWaveBDM}) can be larger
than that of the vacuum Yee wave equation (Eq.~\ref{eq:discWaveB}).
Therefore, the Dey--Mittra time-domain algorithm can require a much
reduced timestep as compared to $\Delta t_{\rm CFL}$
\cite{nieter2009application}.

Alternatives to and modifications of the Dey--Mittra algorithm have been
suggested that try to avoid the unfavorable CFL condition. The
algorithm of \cite{zagorodnov2003uniformly} is complicated, but
retains the second-order
convergence of Dey-Mittra without reduction in timestep (compared
to $\Delta t_{\rm CFL}$) by expanding the stencil for components with
small area fractions (effectively averaging/enlarging the curl of $\bf E$
update for that component). Another simple modification of Dey--Mittra
retains the small boundary stencil {\it and} the vacuum CFL timestep,
but at the expense of second-order
convergence \cite{zagorodnov2007conformal}; however, it achieves low
absolute frequency errors in the numerical tests. A different approach
corrects for the errors induced by stairstepping the boundary
\cite{tornberg2008consistent,engquist2011energy}; however, to ensure
energy conservation (and thus, long-time numerical stability), linear
solves are required to find the operator coefficients on the boundary.
This has been demonstrated only in 2D \cite{engquist2011energy}.

With Dey--Mittra, the usual approach to avoid prohibitively small
time\-steps is to neglect faces with small area fractions, resulting in
a pointy perturbation of the original conducting boundary.
Unfortunately, neglected faces reduce the frequency convergence of the
Dey--Mittra algorithm to first-order \cite{nieter2009application}.
The alternatives mentioned in the previous paragraph perform well
compared to the Dey--Mittra algorithm with neglected faces.

Our new eigensolver allows the keeping of {\it all} cut faces without
degradation of the solution time (that is, given a simulation and
resolution, solution times are the same
whether faces are neglected or not). Given this
ease, the effects of neglecting faces are probed in more detail in this
paper as compared to the results of Ref.~\cite{nieter2009application}.
Namely, in addition to confirming the first-order frequency
convergence induced by neglecting faces, we also highlight the stagnation
of field convergence for fields on the surface of the conducting boundary.

The technique used to neglect small face fractions
in Ref.~\cite{nieter2009application}
analyzes the Gershgorin Circle for every boundary face to minimize
the number ignored. For simplicity, we have implemented a less
sophisticated face-neglecting technique.
We define a minimum area fraction, $a_{\rm min}$; magnetic field
components associated with area fractions less than $a_{\rm min}$ are
set to zero (i.e.~the face is neglected). Electric field components on
edges of the neglected face are also set to zero.

\section{Eigensolving the Yee vacuum wave equation}
\label{sec:eigsolveCC}

Simulations of interest (especially in accelerator component design)
routinely require millions of grid cells to achieve the desired
accuracy. The curl-curl matrix in Eq.~\ref{eq:discWaveB} or
\ref{eq:discWaveBDM} therefore has millions
of rows and cannot be fully diagonalized. Iterative eigensolvers that
search for only a small subset of the solutions are necessary.
As a simplified preamble to our Dey--Mittra eigensolving methods,
we review in this section the currently preferred technique for
eigensolving the discretized Maxwell's equations in vacuum
(c.f.~Eq.~\ref{eq:discWaveB}).
From there, only minor modifications to the technique will be required
to explain our method.
%The underlying eigensolving method relies
%heavily on an inversion of the curl-curl matrix, which turns out to be
%the crux of the technique. Fortunately, multigrid algorithms excel at
%this task, but only after a careful transformation of the curl-curl
%which we also summarize.

\subsection{Krylov-subspace shift and invert eigensolvers}

Krylov-subspace iterative eigensolvers are widely-used and discussed
in detail by \cite{stewart2001matrix,van2002computational}.
In essence, the technique builds and refines a subspace in which
approximations to true eigenpairs lie. The Krylov subspace is formed by
repeatedly applying the eigensystem matrix to an initial (usually
random) vector.

In Krylov-subspace methods, convergence to an eigenpair with
eigenvalue $\lambda_i$ is faster when $\lambda_i$ is at
one extreme of the spectrum and the relative separation of
neighboring eigenvalues is large, i.e.~$|\lambda_i - \lambda_{i+1}| /
|\lambda_{\rm max} - \lambda_{\rm min}|$ \cite{grimes1991shifted}.
The eigenmodes of interest (especially in accelerator component
design) are often the lowest-frequency resonant modes; therefore, the
usual scheme is to shift and invert the operator so that the
eigenvalues of interest are at the top of the spectrum and are
well-separated. Algebraically, for a generic eigensystem ${\rm H} {\bf
x} = \lambda {\bf x}$, one solves the problem
\begin{equation}
({\rm H} - \sigma {\rm I})^{-1} {\bf x} = \frac{1}{\lambda - \sigma}
{\bf x}
\label{eq:shiftInv}
\end{equation}
where I is the identity and $\sigma$ is some shift on the order of the
smallest nonzero eigenvalue of H. An eigenvector of ${\rm H} {\bf
x} = \lambda {\bf x}$ is also an eigenvector of Eq.~\ref{eq:shiftInv}.
The shift is used primarily to avoid
nullspaces of H (which render it uninvertible). Theoretically,
the shift could also accelerate convergence to {\it any} eigenpair
by choosing $\sigma \approx \lambda_i$ if eigenpair
$\lambda_i, {\rm x}_i$ is desired. However, in practice, the inversion of ${\rm
H} - \sigma {\rm I}$ often (and quickly) becomes intractable as $\sigma$
increases toward the interior of the spectrum.

\subsection{Inverting the curl-curl matrix}

Building the Krylov subspace for the shifted and inverted system
involves the repeated application of $({\rm H} - \sigma {\rm
I})^{-1}$.
Because H (in our case, $\rm C C^T$) is large, iterative
linear solvers are required. The iterative inversion of Maxwell's
curl-curl operator is well-studied and the current methods of choice
rely heavily on the multigrid technique
\cite{briggs2000multigrid,vanvek1996algebraic}.  When used on
discretizations of the Laplacian operator in model problems
(e.g.~Poisson's equation on a Cartesian grid on a periodic
square/cubic domain), multigrid methods can reduce residuals by an
order of magnitude per linear solver iteration
\cite{briggs2000multigrid} (for a generic linear system ${\rm H}{\bf
x} = {\bf y}$ and an approximate solution $\tilde{\bf x}$, the residual
is $\| {\rm H}\tilde{\bf x} - {\bf y} \|$).
Modifications of standard multigrid
techniques try to reproduce this behavior for the curl-curl operator
\cite{hiptmair1998multigrid,bochev2004improved,clemens2005construction,arbenz2005multilevel,hu2006toward,kolev2009parallel}.

Special treatment must be given to the curl-curl operator because of
its large nullspace (in the case of Eq.~(\ref{eq:discWaveB}), the
nullspace is the set of all modes with nonzero divergence and $k^2 =
0$ plus the three uniform field modes); the
existence of such a nullspace generally hinders the performance of standard
multigrid preconditioners \cite{hiptmair1998multigrid}.  Since
multigrid preconditioners perform admirably on Laplacian-like matrices, one
approach is to augment the curl--curl operator with a grad--div part to look
more like a vector Laplacian (since $\nabla \times \nabla \times - \nabla \nabla
\cdot = -\nabla^2$) in such a way that the electromagnetic eigenmodes are
unchanged \cite{clemens2005construction,bochev2008algebraic}. This technique
demands that the discretized differential operators satisfy certain vector
identities of continuous space (namely, $\nabla \cdot \nabla \times = 0$ and
$\nabla \times \nabla = 0$) and that the vector Laplacian has a
trivial nullspace (such discretizations are called {\it compatible};
see Ref.~\cite{bochev2006principles}).
Indeed, the standard second-order divergence,
curl, and gradient operators on the Yee mesh meet this requirement.

In vacuum, the Yee magnetic divergence operator, D, acts on $\mathbf{b}$ to
give a cell-centered scalar field $\psi$; for Yee grid cell $ijk$, the
operation is
\begin{equation}
\psi_{ijk} = ({\rm D}{\bf b})_{ijk} = \frac{b_{x|i+1jk} - b_{x|ijk}}{\Delta x} +
  \frac{b_{y|ij+1k} - b_{y|ijk}}{\Delta y} +
  \frac{b_{z|ijk+1} - b_{z|ijk}}{\Delta z}.
\label{eq:divB}
\end{equation}
The Yee layout ensures that $\rm D C = 0$ (the discrete equivalent of
the vector identity, $\nabla \cdot \nabla \times = 0$); also, the
transpose is necessarily zero, which is a discrete version of the
identity, $\nabla \times \nabla = 0$. In fact, the discrete Yee
magnetic gradient operator is $-{\rm D}^{\rm T}$, which, for the
component, $b_{x|ijk}$, takes the form
\begin{equation}
(-{\rm D^T}\psi)_{x|ijk} =
\frac{\psi_{ijk} - \psi_{i-1jk}}{\Delta x}.
\label{eq:gradPhi}
\end{equation}
In the absence of any boundaries, the new eigenproblem to be
solved (which approximates the vector Laplacian) is
\begin{equation}
\left( {\rm C C^{\rm T} + D^{\rm T} D } \right) \mathbf{b}^{\prime} =
{k^{\prime}}^2 \mathbf{b}^{\prime}
\label{eq:discLaplB}
\end{equation}
Because the range of C is in the nullspace of D, an
eigenmode of Eq.~(\ref{eq:discWaveB}) with nonzero eigenvalue
is also an eigenmode of
Eq.~(\ref{eq:discLaplB}) with the same eigenvalue
(i.e.~$k_i^2 = {k_i^{\prime}}^2$ and ${\rm b}_i = {\rm b}_i^{\prime}$
for all $k_i^2 > 0$).
%(to show, simply apply the vector Laplacian
%in Eq.~(\ref{eq:discLaplB}) to an eigenmode of
%Eq.~(\ref{eq:discWaveB})).
%Equation (\ref{eq:discLaplB}) can now be
%solved efficiently with a spectral-shift and invert eigensolver
%using standard multigrid preconditioners.

The form of the matrix in Eq.~\ref{eq:discLaplB} now facilitates
efficient inversion by multigrid methods; however, in exchange,
modes with nonzero divergence (which originally
had eigenvalues equal to zero in Eq.~\ref{eq:discWaveB}) now pollute
the spectrum.

\subsection{Projection}

Modes with nonzero divergence introduced by the grad-div operator
can be removed at any time by the following projection operator
\begin{equation}
\rm P = I - D^T (D D^T)^{-1} D
\label{eq:projYee}
\end{equation}
where the matrix $\rm D D^T$ is a scalar Laplacian and thus
is easily inverted
by multigrid techniques. To see that it is indeed a projection
operator, note that $\rm P^2 = P$.

The projection works because an arbitrary magnetic vector $\bf b$ can
be expressed as the discrete Helmholtz decomposition
\begin{equation}
{\bf b} = {\rm C} {\bf e} + {\rm D^T} \psi
\label{eq:hodge}
\end{equation}
where $\bf e$ is an arbitrary grid electric field (associated with
cell edges) and $\psi$ is an arbitrary cell-centered scalar field.
The two terms are orthogonal under the Euclidean inner product since 
the Yee differential operators
satisfy the vector identities $\rm DC = C^TD^T = 0$. Furthermore, the
two terms span the entire discrete magnetic vector space since it is
known that the vector Laplacian in Eq.~\ref{eq:discLaplB} has a
trivial nullspace (in the absence of constant field vectors; e.g., for
Dirichlet boundary conditions) \cite{bochev2006principles}.

Applying the projection operator to ${\bf b}$ in Eq.~\ref{eq:hodge}
gives
\begin{equation}
{\rm P} {\bf b} = {\rm C} {\bf e}.
\end{equation}

We can now eigensolve Eq.~\ref{eq:discWaveB} for the modes of
interest; the following transformed
eigensystem places eigenvalues of the
low-frequency electromagnetic modes at the top
of the transformed spectrum while zeroing the eigenvalues of the
unwanted modes (with nonzero divergence):
\begin{equation}
{\rm (I - D^T (D D^T)^{-1} D) (C C^T + D^T D - \sigma I)^{-1}}
{\bf b} = \frac{1}{k^2 - \sigma} {\bf b}.
\label{eq:yeeFullMat}
\end{equation}
A single eigensolver iteration requires two matrix
inversions---the shifted vector Laplacian and the scalar Laplacian
within P.

%usually embedded within conjugate-gradient \cite{van1992bi} or
%Krylov-subspace-based linear solvers \cite{saad1986gmres}.

\section{Eigensolving the Dey--Mittra wave equation}
\label{sec:eigsolveDM}

The Dey--Mittra algorithm amounts to a slight modification of the
Yee vacuum curl-curl matrix. Therefore, we hope to find a slight
modification of the vacuum Yee grad-div operator
that complements the modified curl-curl and enables
fast multigrid matrix inversions. The first part of this section
constructs such a grad-div operator. Next, some simple matrix
conditioning is performed to help the inversions and
a projection operator is constructed similar to Eq.~\ref{eq:projYee}.

\subsection{A grad-div for Dey--Mittra}
\label{sec:dmGradDiv}

In this section, we construct a grad-div operator for the Dey--Mittra
algorithm based on physical intuition. Later (Sec.~\ref{sec:results}),
we show that this operator leads to the nearly ideal performance of
standard multigrid preconditioners on the resulting vector Laplacian.
The choice of grad--div is suggested by a discrete form of Gauss' Law
(in integral form) for the magnetic field.

Equation (\ref{eq:divB}) can be rewritten as
\begin{align}
\psi_{ijk} = \frac{1}{v_{ijk}} ( &
  a_{x|i+1jk} b_{x|i+1jk} -
  a_{x|ijk} b_{x|ijk} +
  a_{y|ij+1k} b_{y|ij+1k} - \notag \\
  & a_{y|ijk} b_{y|ijk} +
  a_{z|ijk+1} b_{z|ijk+1} -
  a_{z|ijk} b_{z|ijk} ).
\label{eq:divBInt}
\end{align}
where, in vacuum, $v_{ijk}$ is the volume of Yee grid cell $ijk$ (in
vacuum, $v_{ijk} = \Delta x \Delta y \Delta z$). The above can be
interpreted as a discrete representation of Gauss' Law for the
divergence of $\mathbf{B}$: $\int \nabla \cdot \mathbf{B} dv = \oint
\mathbf{B} \cdot d\mathbf{a}$; $\psi_{ijk}$ represents the value of
the magnetic divergence averaged over the volume of grid cell $ijk$
and is calculated from the total flux of $\mathbf{B}$ out of that grid
cell.

\begin{figure}[htbp]
\centering
\includegraphics[scale=0.4]{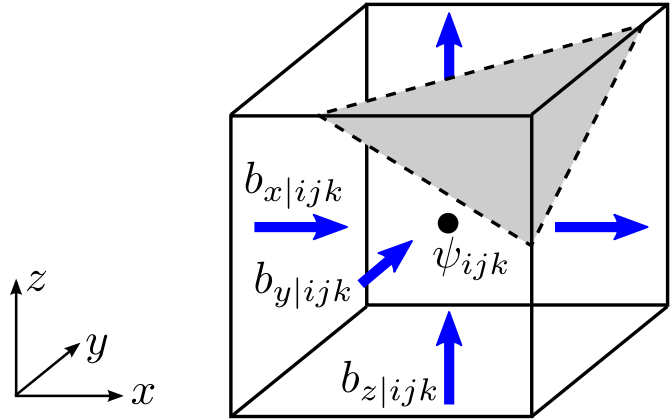}
\caption{An example of cut cell $ijk$. The shaded region is conductor.
We calculate the discrete divergence of $\bf B$ by summing the magnetic
fluxes out of the unshaded volume and dividing by the unshaded volume.
Since the normal magnetic field is zero on the conducting surface,
that outward flux is ignored.}
\label{fig:dmVol}
\end{figure}

Equation (\ref{eq:divBInt}) is naturally extended to Dey--Mittra
boundaries where, for example, $a_{x|ijk} < \Delta y \Delta z$ and
$v_{ijk} < \Delta x \Delta y \Delta z$ (see Fig.~\ref{fig:dmVol} for
an example of a cut cell). If grid cell $ijk$ is cut, $v_{ijk}$ is
the volume of the cell that is {\it outside} the conductor. To find
the volume-averaged divergence, $\psi_{ijk}$, the outward flux of
$\mathbf{B}$ must be calculated on the bounding surfaces of $v_{ijk}$.
Fortunately, the conducting boundary
condition on $\mathbf{B}$, Eq.~(\ref{eq:normB}), forces the normal
component to zero, so that the
boundary surface is not included in the flux calculation, leaving
Eq.~(\ref{eq:divBInt}) a physically meaningful expression.
%In matrix-vector form, the
%Dey--Mittra divergence operator gives
In matrix form, the Dey--Mittra divergence operator is
\begin{equation}
%\phi = {\rm V^{-1} D A } \mathbf{b}
{\rm D}_{DM} = {\rm V^{-1} D A }
\label{divBDM}
\end{equation}
where V is a diagonal matrix of cell volume fractions
(${\rm V}_{(ijk)(ijk)} = v_{ijk} / \Delta x \Delta y \Delta z$).

With Dey--Mittra boundaries, the new eigenproblem for the discrete vector Laplacian is then
\begin{equation}
\left( {\rm A^{-1} C L C^{\rm T} + D^{\rm T} V^{-1} D A } \right)
\mathbf{b}^{\prime} = {k^{\prime}}^2 \mathbf{b}^{\prime}.
\label{eq:discLaplBDM}
\end{equation}
Conveniently, the modified Dey--Mittra divergence operator has also guaranteed that
eigenmodes of Eq.~(\ref{eq:discWaveBDM}) with $k^2 \neq 0$
(electromagnetic eigenmodes) are also eigenmodes of
Eq.~(\ref{eq:discLaplBDM}) since ${\rm D A} \mathbf{b} = 0$ when
$\mathbf{b}$ is an eigenmode of
Eq.~(\ref{eq:discWaveBDM}); in fact, the
inverse volume fraction matrix is irrelevant for this purpose---setting
it equal to the identity would also ensure that electromagnetic
eigenmodes are unchanged from Eq.~(\ref{eq:discWaveBDM}) to Eq.~(\ref{eq:discLaplBDM}).
However, as we will see in the numerical
results section, the accurate calculation of the volume fractions is
essential for good performance of multigrid preconditioners.

\subsection{Conditioning and projecting}
\label{sec:eigsolver}

%When only a tiny fraction of a grid cell face that is cut by a
%conducting boundary is outside the conductor, it is possible that
%a row of $\rm A^{-1} C L C^{\rm T}$ contains very large values
%(matrix elements are proportional to $l/a \sim 1/l$ for general edge
%fractions $l$ and face fractions $a$, and can get large
%for $l \rightarrow 0$). The same goes for the matrix $\rm D^T V^{-1} D
%A$ and small volume fractions.
%By the Gershgorin Circle theorem, these large
%elements inflate the upper eigenvalue bound of the Dey--Mittra curl-curl matrix
%and therefore decrease the
%CFL limit on the timestep when using the Dey--Mittra
%algorithm in the time-domain \cite{nieter2009application}.

We have found it quite difficult to invert the left-hand side of
Eq.~\ref{eq:discLaplBDM}; however, if we multiply both sides by the
area fraction matrix, A, then inversions using multigrid converge
very quickly. We believe this is primarily due to
ill-conditioning of the matrix in Eq.~\ref{eq:discLaplBDM} resulting
from large elements in ${\rm A}^{-1}$ and ${\rm V}^{-1}$. 
Applying the spectral shift and inversion to the A-transformed system
gives
\begin{equation}
\left( {\rm C L C^{\rm T} + A D^{\rm T} V^{-1} D A} - \sigma {\rm A} \right)^{-1} {\rm A}
\mathbf{b}^{\prime} = \frac{1}{{k^{\prime}}^2 - \sigma}
\mathbf{b}^{\prime}.
\label{eq:discLaplDM}
\end{equation}
It was also
noted by one of our reviewers that the matrix in
Eq.~\ref{eq:discLaplDM} is
symmetric, which may further help the linear solver.

%this issue is eliminated, resulting in
%the following
%\begin{equation}
%\left( {\rm C L C^{\rm T} + A D^{\rm T} V^{-1} D A } \right)
%\mathbf{b}^{\prime} = {k^{\prime}}^2 {\rm A} \mathbf{b}^{\prime}.
%\label{eq:discLaplBDMcond}
%\end{equation}
%This
%general ill-conditioning of the matrix in Eq.~\ref{eq:discLaplBDM}
%(the grad--div operator is also ill-conditioned since
%$a/v \sim 1/l$ for general volume fractions $v$)
%also hinders the performance of iterative linear solvers, which are
%of primary importance in spectral-shift and invert eigensolvers.
%Fortunately, we can easily rewrite Eq.~\ref{eq:discLaplBDM} as
%since A is just a diagonal matrix. The left matrix is now
%better-conditioned (boundary elements scale as $l$ in the
%curl--curl part and as $a^2/v \sim l$ in the grad--div part) and
%symmetric.
%Applying the spectral shift and invert then gives
%\begin{equation}
%\left( {\rm C L C^{\rm T} + A D^{\rm T} V^{-1} D A} - \sigma {\rm A} \right)^{-1} {\rm A}
%\mathbf{b}^{\prime} = \frac{1}{{k^{\prime}}^2 - \sigma}
%\mathbf{b}^{\prime}.
%\label{eq:discLaplDM}
%\end{equation}

As with the Yee scheme in vacuum, the eigenmodes with nonzero
divergence
must be removed (post-inversion) from the range of the
vector Laplacian.
Let the inner product on the discrete Yee magnetic field be
\begin{equation}
(\mathbf{b}_1^{\prime}, \mathbf{b}_2^{\prime}) \equiv
{\mathbf{b}_1^{\prime}}^{\rm T} {\rm A}
\mathbf{b}_2^{\prime}.
\label{eq:bInner}
\end{equation}
then we can write an arbitrary discrete magnetic field vector
$\mathbf{b}^{\prime}$, as the Helmholtz decomposition
\begin{equation}
\mathbf{b}^{\prime} = {\rm A^{-1} C L} \mathbf{e} + {\rm D}^{\rm T} \psi
\end{equation}
where $\mathbf{e}$ is a discrete Yee electric field vector and $\psi$ is some
cell-centered scalar field (the two parts
are orthogonal under the inner product of Eq.~\ref{eq:bInner}).
For a general vector, $\mathbf{b}^{\prime}$, with the above
decomposition, the following projection operator will eliminate the
part derived from the gradient matrix:
\begin{equation}
{\rm P}_{DM} \equiv {\rm I - D^T ( D A D^T )^{-1} D A }
\label{eq:proj}
\end{equation}
where I is the identity matrix. In all of our numerical tests,
the matrix $\rm D A D^T$ was easily inverted by multigrid.
Generally, we have found that if the linear solver
can invert the vector Laplacian, it can invert the scalar Laplacian
in Eq.~\ref{eq:proj} in a shorter time and in fewer iterations.

The final transformed system to be diagonalized iteratively is
\begin{equation}
\left({\rm I - D^T ( D A D^T )^{-1} D A }\right)
\left( {\rm C L C^{\rm T} + A D^{\rm T} V^{-1} D A} - \sigma {\rm A} \right)^{-1} {\rm A}
\mathbf{b}^{\prime} = \frac{1}{{k^{\prime}}^2 - \sigma} \mathbf{b}^{\prime}
\label{eq:fullMat}
\end{equation}
The largest eigenvalues of the above matrix now correspond to
${k^{\prime}}^2 \approx \sigma$, are relatively well-separated,
and are purely electromagnetic
(since the projection operator zeros the eigenvalues of modes with
nonzero divergence).

\subsection{A transformation to resemble Yee}

It was observed by one of our reviewers that
Eq.~\ref{eq:fullMat}
can be written in a more compact form that makes a stronger connection
with the Yee scheme. If we let
\begin{equation}
\tilde{\rm D} = {\rm V}^{-1/2} {\rm D A}^{1/2} \quad
\tilde{\rm C} = {\rm A}^{-1/2} {\rm C L}^{1/2} \quad
\tilde{\bf b} = {\rm A}^{1/2} {\bf b}
\label{eq:transformation}
\end{equation}
then left-multiplying Eq.~\ref{eq:fullMat} by ${\rm A}^{1/2}$ gives
\begin{equation}
\left({\rm I - \tilde{D}^T ( \tilde{D} \tilde{D}^T )^{-1} \tilde{D}}\right)
\left( {\rm \tilde{C} \tilde{C}^T + \tilde{D}^T \tilde{D}} - \sigma {\rm I} \right)^{-1}
\tilde{\mathbf{b}}^{\prime} = \frac{1}{{k^{\prime}}^2 - \sigma}
\tilde{\mathbf{b}}^{\prime}
\label{eq:fullMat2}
\end{equation}
which mimics Eq.~\ref{eq:yeeFullMat}. The transformations of
Eq.~\ref{eq:transformation} also compactly
represent the vector identities and Helmholtz decomposition for the
Dey--Mittra algorithm:
\begin{align}
{\rm \tilde{D} \tilde{C}} &= {\rm \tilde{C}^T \tilde{D}^T} = 0 \\
\tilde{\bf b}^{\prime} &= {\rm \tilde{C}} {\bf e} + {\rm \tilde{D}^T}
\psi,
\end{align}
respectively.

Unfortunately, numerical tests show that the elegant symmetric form of
Eq.~\ref{eq:fullMat2} ruins convergence, supporting the argument
that the inflated elements of the 
${\rm A}^{-1/2}$ operator within $\tilde{C}$ are to blame (rather than
overall operator symmetry). The form in Eq.~\ref{eq:fullMat} was
used to obtain the following numerical results.

\section{Numerical results}
\label{sec:results}

Our eigensolver makes extensive use of the \textsc{trilinos} linear
algebra framework \cite{trilinos} to solve Eq.~\ref{eq:fullMat}.
Specifically, we use the block
Krylov-Schur routine from the Anasazi package for the outer
eigensolver iterations \cite{stewart2002krylov,baker2009anasazi}, the
GMRES linear solver from the AztecOO package for matrix inversions
\cite{saad1986gmres}, and the algebraic multigrid (AMG) tool from the
ML package as a preconditioner for the GMRES solver
\cite{vanvek1996algebraic,ml-guide}. Within the AMG method, we use the
polynomial-based multilevel smoother (of order 1) as described in
\cite{adams2003parallel} which exhibits good parallel performance
(e.g.~as compared with popular Gauss-Seidel smoothers) since only
matrix-vector multiplications are required.
The multigrid preconditioner used the V-cycle; therefore, the smoother
was applied twice at each level per iteration (once on the way to
coarser grids, and once on the way back to the original fine grid).
The coarsest level was treated the same as any other multigrid level.
The vector and scalar Laplacians to be inverted in
Eq.~\ref{eq:fullMat} were formed explicitly, then passed to the ML
algorithm.

Cubic grid cells were used throughout the following tests ($\Delta x =
\Delta y = \Delta z$).

Most simulations were performed on Hopper, the Cray XE6 supercomputer
at the National Energy Research Scientific Computing Center.

%\subsection{Multigrid/Linear solver performance}
\subsection{Performance: spherical cavity}

In Table \ref{tab:invPerf}, we compare the performance of the linear
solver (the bottleneck) on a model problem
(cubic domain with perfectly-conducting boundaries) with a
simple problem requiring the
Dey--Mittra algorithm (the spherical cavity). No cut faces were
neglected
in the sphere simulation; the smallest area fraction encountered was
$6 \times 10^{-4}$. The cubic
domain had arbitrary side-length $L$, and the spherical cavity had
radius $0.49 L$. Eigensolves were performed for the lowest three
modes; each eigensolve took about 20 outer iterations to complete (20
vector/scalar Laplacian inversions). The figures in Table
\ref{tab:invPerf} are averages over these 20 inversions.

The multigrid complexity is defined as
\begin{equation}
{\rm complexity} = \frac{\sum_{i=1}^N
\operatorname{nnz}({\rm A}_i)}{\operatorname{nnz}({\rm A}_1)}
\end{equation}
where A$_i$ is the linear system matrix on multigrid level $i$ (with
$i = 1$ representing the finest level) and the nnz() operator produces
the number of nonzero elements in the operand matrix.

%A higher
%complexity indicates a more expensive multigrid routine. Complexity is
%a result of the aggregation algorithm. For discussion of a more
%careful aggregation strategy on the Yee mesh,
%see \cite{bochev2008algebraic}.
%The multigrid complexities reported in Tab.~\ref{tab:invPerf} induced
%by the ``uncoupled'' aggregation scheme in ML are high compared to
%those achieved by Ref.~\cite{bochev2008algebraic}. The specialized
%aggregation scheme described in Ref.~\cite{bochev2008algebraic} may help,
%but has not yet been implemented.

\begin{table}[h]
\footnotesize
\centering
\caption{Linear solver benchmarks for the inversion of the
vector Laplacian without (cube) and with (sphere) Dey--Mittra boundaries.
One polynomial (order 1) smoother sweep per multigrid level.
{\it Component count} is the number of magnetic field components.
Inversions were performed to $10^{-6}$ accuracy. For every simulation,
processor domains were 16x16x16 cells.}
%\begin{tabular}{|*{4}{c|}}
\begin{tabular}{|r||c|c|c||c|c|c||}
\hline
& \multicolumn{3}{|c||}{Cube} & \multicolumn{3}{|c||}{Sphere} \\ \hline \hline
%Component count & 11,520 & 95,232 & 774,144 & 7,224 & 53,160 & 405,876 \\ \hline
%Avg.~iteration count & 8.0 & 8.5 & 9.6 & 7.0 & 8.0 & 8.5 \\ \hline
%Convergence rate & 0.18 & 0.20 & 0.24 & 0.14 & 0.18 & 0.20 \\ \hline
%%Avg.~time/inversion (ms) & 46 & 128 & 396 & 28 & 64 & 292 \\ \hline
%Multigrid levels & 4 & 4 & 4 & 3 & 4 & 4 \\ \hline
%Multigrid complexity & 4.1 & 4.6 & 4.9 & 3.6 & 4.5 & 4.9 \\ \hline
%Domain decomposition & 1x1x1 & 2x2x2 & 4x4x4 & 1x1x1 & 2x2x2 & 4x4x4 \\ \hline
Component count & 12,285 & 98,301 & 786,429 & 7,815 & 55,398 & 414,489 \\ \hline
Avg.~iteration count & 8.8 & 9.3 & 9.6 & 10.0 & 10.1 & 10.5 \\ \hline
Convergence rate & 0.21 & 0.22 & 0.24 & 0.25 & 0.25 & 0.27 \\ \hline
%Avg.~time/inversion (ms) & 46 & 128 & 396 & 28 & 64 & 292 \\ \hline
Multigrid levels & 3 & 4 & 4 & 3 & 4 & 4 \\ \hline
Multigrid complexity & 1.5 & 1.6 & 1.6 & 1.6 & 1.7 & 1.8 \\ \hline
Domain decomposition & 1x1x1 & 2x2x2 & 4x4x4 & 1x1x1 & 2x2x2 & 4x4x4 \\ \hline
\end{tabular}
\label{tab:invPerf}
\end{table}

Iteration counts are strikingly similar between the model problem and
the spherical cavity.
However, if the inverse volume matrix is left out of
Eq.~\ref{eq:fullMat}, vector Laplacian inversions do not converge in a
reasonable number of iterations (for a GMRES basis size of 40, more
than 10 restarts were required---thus, more than 400 total iterations).
To examine the importance of the inverse volume matrix on the
linear solver performance, we perturbed our accurate cell volume
calculations (for cut cells only) in the following way
\begin{equation}
\tilde{v}_{ijk} = 10^{f_{ijk} \epsilon} v_{ijk}
\label{eq:volPert1}
\end{equation}
where $v_{ijk}$ is the accurate volume, $f_{ijk}$ is a random factor
between $-1$ and 1, and $\epsilon$ is an ``order of magnitude''
parameter (e.g.~for $\epsilon=1$, the volume can be wrong by up to an
order of magnitude).

We also tested errors that are likely to result
from a simple subsampling volume calculation routine. In this case,
the model was
\begin{equation}
\tilde{v}_{ijk} = v_{ijk} + f_{ijk} \Delta v
\label{eq:volPert2}
\end{equation}
where $v_{ijk}$ is the accurate volume, $f_{ijk}$ is a random factor
between $-1$ and 1, and $\Delta v$ is the perturbation (or subsample)
volume.

Table \ref{tab:volPert1} shows the effect of these errors on linear
solver convergence for different values of $\epsilon$ and $\Delta v /
v_{\rm vac}$ where $v_{\rm vac} = \Delta x \Delta y \Delta z$, and
indicates the importance of a good volume estimate (assuming accurate
area fraction calculations).  For example, according to Table
\ref{tab:volPert1}, a subsampling routine with 10 samples per
dimension
of a grid cell (1000 cubic subvolumes) could lead to
very poor convergence.  Our
volume calculation method breaks cut-cell volumes into tetrahedral
subvolumes that conform to the embedded surface
(the volumes of which are easy to calculate individually);
therefore, our method converges to the exact cut-cell volume for
smooth boundaries.

%\ref{tab:volPert2} shows the effect of these errors on linear solver
%convergence for different values of $\Delta v / v_{\rm vac}$ where
%$v_{\rm vac} = \Delta x \Delta y \Delta z$. According to Table
%\ref{tab:volPert},
%a simple subsampling routine for calculating cut-cell volumes would be
%quite inadequate at
%10x10x10 subsamples---already a computational burden when applied to
%many cut
%cells. Our method used a tetrahedral filling of each cut cell, and
%converges to the exact cut cell volume for smooth boundaries.

\begin{table}[h]
\centering
\caption{Effect of cut-cell volume calculation errors on linear solver
convergence for the spherical cavity problem.  Each $\epsilon$ and
$\Delta v$ was
simulated three times; the average iteration count from each was then
averaged over these three
simulations.
Simulation domain was
32x32x32. Inversions were performed to $10^{-6}$ accuracy.}
%\begin{tabular}{|*{4}{c|}}
\begin{tabular}{|c|c|c|c|c|c|c|}
\hline
\multicolumn{7}{|c|}{Error from Eq.~\ref{eq:volPert1}} \\ \hline
$\epsilon$ & 0.1 & 0.3 & 0.5 & 1 & 2 & 3 \\ \hline
Avg.~iteration count & 10 & 11 & 13 & 24 & 74 & 220 \\ \hline \hline
\multicolumn{7}{|c|}{Error from Eq.~\ref{eq:volPert2}} \\ \hline
$\Delta v / v_{\rm vac}$ &
\multicolumn{2}{|c}{$10^{-5}$} &
\multicolumn{2}{|c}{$10^{-4}$} &
\multicolumn{2}{|c|}{$10^{-3}$} \\ \hline
Avg.~iteration count &
\multicolumn{2}{|c}{10} &
\multicolumn{2}{|c}{50} &
\multicolumn{2}{|c|}{91} \\ \hline
\end{tabular}
\label{tab:volPert1}
\end{table}

%\begin{table}
%\centering
%\caption{Effect of cut-cell volume calculation errors on linear solver
%convergence for the spherical cavity problem. Simulation domain was
%32x32x32.
%Inversions were performed to $10^{-6}$ accuracy.}
%%\begin{tabular}{|*{4}{c|}}
%\begin{tabular}{|c|c|c|c|}
%\hline
%$\Delta v / v_{\rm vac}$ & $10^{-5}$ & $10^{-4}$ &
%$10^{-3}$ \\ \hline
%Ave.~iteration count & 8 & 37 & 200 \\ \hline
%\end{tabular}
%\label{tab:volPert}
%\end{table}

\subsection{Dey--Mittra frequency and field convergence}
\label{sec:conv}

This section compares frequencies and surface fields in the spherical
cavity calculated by the Dey--Mittra algorithm with their analytic
counterparts for varying $a_{\rm min} / a_{\rm vac}$ (where
$a_{\rm
vac} = \Delta x \Delta y = \Delta y \Delta z = \Delta x \Delta z$
because cells are cubic). For
these simulations, parity symmetry was invoked to reduce the
simulation domain to the positive $x$, $y$, and $z$ octant (sphere
centered at the origin). Boundaries at $x=0$, $y=0$, and $z=0$
(the symmetry planes) were set as perfectly-conducting. The lowest
eigenmode was analyzed which, for the above symmetry, corresponded to
the TM$_{032}$ spherical cavity mode \cite{jackson1999ced}.

\begin{figure}[htbp]
\centering
\includegraphics[scale=0.6]{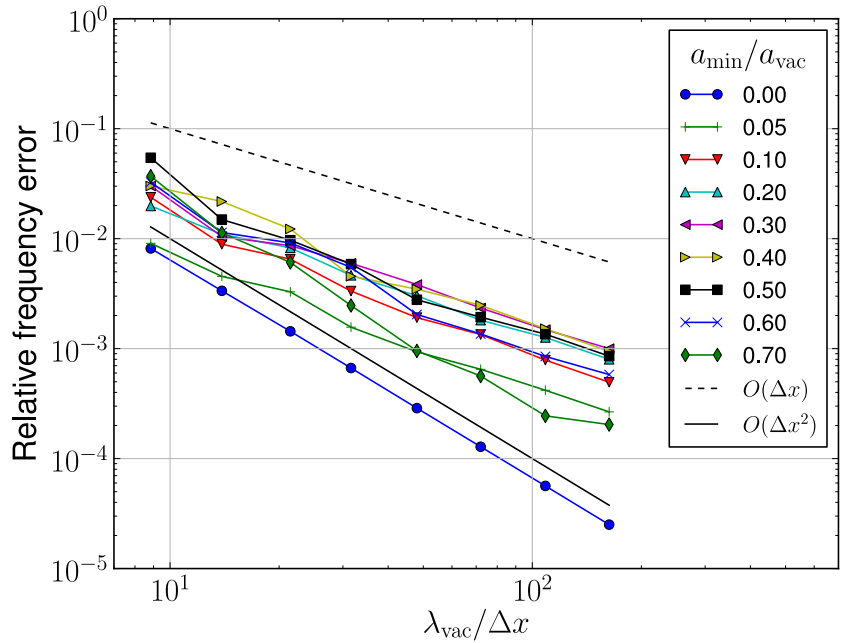}
\caption{Color online. Convergence of TM$_{032}$ resonant frequency
in the spherical cavity as a function of resolution and
$a_{\rm min} / a_{\rm vac}$. The
unaltered Dey--Mittra algorithm is clearly second-order, while
neglecting faces results in eventual first-order convergence (as
similarly reported in \cite{nieter2009application})}
\label{fig:freqErr3d}
\end{figure}

Relative frequency errors were calculated by
\begin{equation}
\varepsilon_{\rm freq} = \frac{|\omega - \omega_0|}{\omega_0}
\end{equation}
where $\omega$ is the calculated angular frequency and $\omega_0$
is the analytic angular frequency. All plots in this section use as
the abscissa the resolution relative to the vacuum wavelength of the
mode, where the vacuum
wavelength is defined as $\lambda_{\rm vac} = 2 \pi c / \omega_0$.
Figure \ref{fig:freqErr3d} shows the second-order convergence of the
pure Dey--Mittra algorithm and the first-order error effect introduced
by neglecting cut faces. One should also note the difference in relative
error at low resolutions, which can be significant.

Electric fields near the surface of the spherical cavity were
calculated and compared with analytic solutions
\cite{jackson1999ced}.  Since interpolation of fields to the metal
boundary is equivocal and nontrivial, we analyzed ``surface'' fields
on a shell a
radial distance $3 \Delta x$ away from the boundary where standard
trilinear interpolations are appropriate.
Because surface fields were analyzed a fixed number of grid cell
widths from the boundary, the {\it physical} distance from the boundary
decreased as resolution increased. Before
surface fields were compared, analytic and simulated fields were
scaled to have the same $\ell_2$-norm where the norm was calculated by
sampling each field over the same collection of points filling the
cavity volume.

\begin{figure}[htbp]
\centering
\includegraphics[scale=0.6]{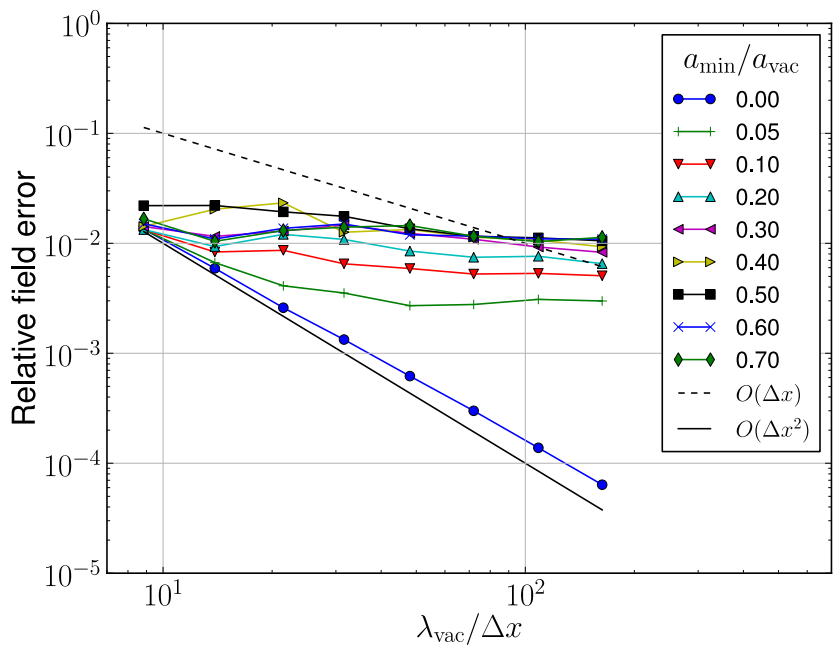}
\caption{Color online. Convergence of TM$_{032}$ surface fields
in the spherical cavity as a function of resolution and
$a_{\rm min} / a_{\rm vac}$.}
\label{fig:fieldErr3d}
\end{figure}

The relative $\ell_2$ errors of the computed eigenmodes are,
\begin{equation}
\varepsilon_{\rm field} = \sqrt{\frac{\sum_i \big| {\bf e}({\bf x}_i) -
{\bf E}({\bf x}_i) \big| ^2}{\sum_i \big| {\bf E}({\bf x}_i) \big|^2}}
\label{eq:errorNorm}
\end{equation}
where the ${\bf x}_i$ are test points on a shell,
${\bf E}({\bf x})$ is the analytic eigenmode evaluated
at $\bf x$, and ${\bf e}({\bf x})$ is the
computed eigenmode interpolated to the point $\bf x$ (using trilinear
interpolation).
Figure \ref{fig:fieldErr3d} shows the results for the TM$_{032}$
spherical cavity mode. Surface fields ultimately do not converge when
faces are neglected, yet converge with nearly second-order error
for the pure Dey--Mittra algorithm.

\subsection{Performance: TESLA cavity}

\begin{figure}[htbp]
\centering
\includegraphics[scale=0.5]{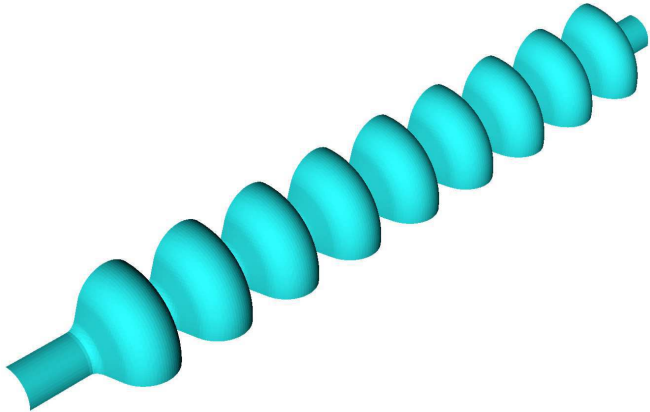}
\caption{Simulated quarter section of the 9-cell TESLA
superconducting accelerator cavity. Beam tubes terminate at
conducting walls. The length of the cavity is 1.38 meters.}
\label{fig:tesla}
\end{figure}

Our eigensolver's performance (i.e.~the linear solver
performance) was also tested on a modern
accelerator cavity problem---the TESLA
superconducting cavity \cite{aune2000super} (see
Fig.~\ref{fig:tesla}). One quarter of the nine-cell cavity
was simulated where the $x=0$ and $y=0$ boundary planes were
perfect magnetic conductor (the $z$-direction is the long direction);
therefore, the TM$_{010}$-like
accelerating modes were found at the lowest frequencies ($\approx 1.3$
GHz). Three different resolutions were tested on the TESLA cavity
shown in Fig.~\ref{fig:tesla}; additionally, a {\it cryomodule} was
simulated, which consists of 8 TESLA cavities strung end-to-end in the
$z$-direction.

Simulation results are summarized in Table
\ref{tab:teslaPerf}; the first 3 columns refer to the single TESLA
cavity simulations and the last column the cryomodule.
For the TESLA cavity simulations,
the maximum eigensolver Krylov basis size (before
a restart occurs) was 50, and the lowest 9 eigenmodes were
found. Only 32 outer eigensolver iterations were required
(no restarts necessary) at each resolution. For the cryomodule
simulation, to save resources on Hopper (supercomputer at NERSC), we
restricted the eigensolver to find only the lowest 3 eigenmodes
and disallowed eigensolver restarts; 30 eigensolver iterations were
performed.

In Table \ref{tab:teslaPerf}, the estimated relative frequency error
of the accelerating mode is given by $\varepsilon_{\rm freq} = (f \Delta
x / c)^2$ where $f = 1.3$ GHz.

\begin{table}[h]
\centering
\caption{Performance of linear solver during an eigensolve
of the TESLA cavity. One polynomial (order 1) smoother sweep per level per
traverse (same as spherical cavity simulations). Inversions
performed to $10^{-6}$ accuracy. Numbers in the last column apply to a
cryomodule simulation---eight TESLA cavities end-to-end.}
\begin{tabular}{|r||c|c|c|c|}
\hline
Comp.~count ($\times 10^6$) & 0.8 & 6.4 & 50.2 & 520 (cryo) \\ \hline
Grid cell $\Delta x$ (mm) & 2.9 & 1.4 & 0.72 & 0.66 \\ \hline
Estimated $\varepsilon_{\rm freq}$ & $2 \times 10^{-4}$ &
  $4 \times 10^{-5}$ &
  $1 \times 10^{-5}$ &
  $8 \times 10^{-6}$ \\ \hline
Avg.~iteration count & 11.8 & 11.9 & 12.6 & 13.9 \\ \hline
Convergence rate & 0.31 & 0.31 & 0.33 & 0.37 \\ \hline
Multigrid complexity & 1.6 & 1.6 & 1.6 & 1.6 \\ \hline
Multigrid levels & 5 & 5 & 5 & 5 \\ \hline
%Avg.~time/inversion (s) & 1.9 \\ \hline
Domain decomposition & 1x1x12 & 2x2x24 & 4x4x48 & 4x4x418 \\ \hline
\end{tabular}
\label{tab:teslaPerf}
\end{table}

\section{Conclusion}

We derived a matrix transformation for the Dey--Mittra
curl-curl operator that allows for efficient inversion by
multigrid methods.  In fact, the performance of multigrid
applied to our custom
Dey--Mittra vector Laplacian nearly equals the performance of
multigrid on the model problem (grid-aligned cubic domain).
As a result,
we developed an efficient shift-and-invert eigensolver for
Maxwell's equations in resonant cavities.  Eigensolver
performance was demonstrated on the simple spherical cavity
and the TESLA superconducting cavity. This eigensolver is
open-sourced at \url{https://github.com/bauerca/maxwell}.

The effects on accuracy of neglecting small cut faces in the
Dey--Mittra algorithm were also investigated. An analysis of
Dey--Mittra surface fields showed the stagnation of
convergence (when faces are neglected) for fields a fixed
number of grid cells away from conducting boundaries; in
contrast, if all faces are kept (which is the case for our
new eigensolver), surface field convergence is nearly
second-order.

\section*{Acknowledgement}

This research used resources of the National Energy Research
Scientific Computing Center, which is supported by the Office of
Science of the U.S. Department of Energy under Contract No.
DE-AC02-05CH11231. The authors would also like to thank the reviewers
for their careful read of our manuscript and the Trilinos
team for their mailing-list support.

\bibliographystyle{elsarticle-num}
\bibliography{dmMultigrid}

\end{document}